\documentclass[conference]{IEEEtran}
\IEEEoverridecommandlockouts

\usepackage{cite}
\usepackage{amsmath,amssymb,amsfonts}
\usepackage{algorithmic}
\usepackage{graphicx}
\usepackage{textcomp}
\usepackage{xcolor}
\def\BibTeX{{\rm B\kern-.05em{\sc i\kern-.025em b}\kern-.08em
    T\kern-.1667em\lower.7ex\hbox{E}\kern-.125emX}}
\begin{document}

\title{Robust Intrusion Detection System with Explainable Artificial Intelligence
 \\
{\footnotesize \textsuperscript{}}
\thanks{This work was funded by The Scientific and Technological Research Council of Turkey, under 1515 Frontier R\&D Laboratories Support Program with project (Grant number: 5169902), and in part by the ROBUST-6G Project through the Smart Networks and Services Joint Undertaking (SNS JU) under the European Union’s Horizon Europe Research and Innovation Program (Grant number: 101139068).}
}

\author{\IEEEauthorblockN{1\textsuperscript{st} Bet\"{u}l G\"{u}ven\c{c} Paltun}
\IEEEauthorblockA{\textit{Ericsson Research} \\
Istanbul, Turkey \\
betul.guvenc.paltun@ericsson.com}
\and
\IEEEauthorblockN{2\textsuperscript{nd} Ramin Fuladi}
\IEEEauthorblockA{\textit{Ericsson Research} \\
Istanbul, Turkey \\
ramin.fuladi@ericsson.com}
\and
\IEEEauthorblockN{3\textsuperscript{rd} Rim El Malki}
\IEEEauthorblockA{\textit{Ericsson Standards and Technology} \\
Paris, France \\
rim.el.malki@ericsson.com}
}

\maketitle

\begin{abstract}

Machine learning (ML) models serve as powerful tools for threat detection and mitigation; however, they also introduce potential new risks. Adversarial input can exploit these models through standard interfaces, thus creating new attack pathways that threaten critical network operations. As ML advancements progress, adversarial strategies become more advanced, and conventional defenses such as adversarial training are costly in computational terms and often fail to provide real-time detection. These methods typically require a balance between robustness and model performance, which presents challenges for applications that demand instant response. To further investigate this vulnerability, we suggest a novel strategy for detecting and mitigating adversarial attacks using eXplainable Artificial Intelligence (XAI). This approach is evaluated in real time within intrusion detection systems (IDS), leading to the development of a zero-touch mitigation strategy. Additionally, we explore various scenarios in the Radio Resource Control (RRC) layer within the Open Radio Access Network (O-RAN) framework, emphasizing the critical need for enhanced mitigation techniques to strengthen IDS defenses against advanced threats and implement a zero-touch mitigation solution. Extensive testing across different scenarios in the RRC layer of the O-RAN infrastructure validates the ability of the framework to detect and counteract integrated RRC-layer attacks when paired with adversarial strategies, emphasizing the essential need for robust defensive mechanisms to strengthen IDS against complex threats.

\end{abstract}


\section{Introduction}

Maintaining the confidentiality, integrity, and availability of Artificial Intelligence (AI) models and their data is critically important. Threats appear in various forms, including adversarial attacks on AI models, data manipulation, and unauthorized access to confidential data\cite{chakraborty2018adversarial}. To mitigate these risks, security strategies involve deploying strong authentication processes, encrypting data both in transit and at rest, and continuously monitoring for anomalies to safeguard AI elements within network architectures. Addressing cybersecurity risks posed by adversarial AI methods remains a significant challenge. Meanwhile, security measures are increasingly reliant on AI/ML models to identify and counteract evolving, sophisticated threats. For example, IDSs are critical for analyzing network activities and detecting suspicious actions that may indicate possible attacks. Advances in ML have improved the IDS functionality, allowing them to identify anomalies more effectively. Their effectiveness is particularly notable given the large volume of data processed in 5G and beyond-mobile network environments. However, attackers continuously seek to diminish the efficiency of ML-based IDSs, thereby intensifying the impact of network attacks such as RRC signaling storms. For example, adversarial techniques can enable adversaries to bypass AI-powered IDS and gain unauthorized access to essential systems or data. To address this increasing threat, we intend to propose an agnostic robust strategy to detect and mitigate such attacks using XAI for real-time evaluation.

To further explore the vulnerability mentioned above, this paper outlines our significant contribution to developing a comprehensive framework focused on identifying and countering adversarial threats in IDSs. We present an agnostic methodology that utilizes XAI techniques to evaluate how adversarial samples impact the interpretations of machine learning models. In addition, we have developed a zero-touch detection strategy specifically aimed at improving IDS capabilities to strengthen defenses against attackers using network-related attacks and other advanced techniques to evade detection mechanisms. By proactively addressing these security gaps, we aim to enhance both the security and resilience of IDSs. Our approach enables IDS not only to detect emerging attack vectors, but also to respond immediately to potential threats even when the number of features is limited. This proactive approach ensures that security measures are effective and flexible, protecting confidential data from adversary actions. As a test setup, the O-RAN infrastructure has been chosen since it focuses heavily on the use of AI/ML to optimize the functionality and capabilities of the network \cite{niknam2022intelligent}. O-RAN has incorporated 3rd Generation Partnership Project (3GPP) standards to enhance its interfaces and protocols, addressing the expanded attack surface that comes with virtualization, open interfaces, and multi-vendor settings. Despite these improvements, the architecture is insufficient to address the specific vulnerabilities of ML/AI capabilities within O-RAN components, such as the Non-Real-Time RAN Intelligent Controller (Non-RT RIC) and the Real-Time RAN Intelligent Controller (Near-RT RIC) \cite{illiano2015detecting}.

\noindent
Our main contributions are as follows. 
\begin{itemize}
    \item Developing a novel approach by incorporating an XAI feature into the ML-based detection framework for real-time assessment. 
    \item The detection model gains deeper insights into the importance of features, leading to better selection of relevant attributes to identify adversarial attacks and reduce false positives.
    \item The idea of XAI feature centers around understanding the behavior of the unseen data, ensuring that its distribution aligns with the standard behavior range observed in training data. The essential part is to assess the distribution of SHAP values within the training data, rather than the training data itself, which was not previously proposed.
    \item The proposed XAI-integrated ML-based detection approach improves the overall effectiveness of the detection mechanism, ensuring better adaptability to evolving attack patterns. 
    \item We introduce a straightforward but impactful mitigation technique that relies on anticipated adversarial examples.
    \item In addition to offering detection and mitigation capabilities, this approach introduces a zero-touch strategy specifically designed to enhance IDS capabilities, thus strengthening defenses against adversaries.
\end{itemize}

\section{Background}

This section provides essential background on the proposed framework. We start with an overview of adversarial AI, followed by an introduction to XAI. Lastly, we give a concise description of RRC signaling storm attacks within O-RAN, which we consider as a use case study in this paper.

\subsection{Adversarial AI}\label{adversarial}
Adversarial AI is the field of AI that focuses on investigating how adversaries might take advantage of AI systems by fooling them into providing incorrect results. These modifications, known as adversarial attacks, can lead machine learning models to produce inaccurate predictions or classifications by gradually changing the input data in a way that is often imperceptible to humans but has significant impacts on how well the model is able to make decisions. Attackers may, for example, craft inputs that evade IDS, spam filters, or malware detection algorithms, causing them to mistakenly identify malicious activity as normal or vice versa. It is possible to weaken AI systems using well-crafted adversarial examples, which require strong defenses and a thorough understanding of how these attacks operate.
Adversarial attacks can be broadly divided into white-box and black-box attacks. In white-box attacks, the attacker has complete information about the target model, including its architecture, parameters, and training information. This information allows an attacker to create very effective countermeasures. Black-box attacks, on the other hand, assume that the attacker has access only to the inputs and outputs and is not informed about the inner workings of the model.

\subsection{Explainable AI}\label{xai}

XAI is an important subfield of AI and addresses the need for transparency in AI systems by explaining internal mechanisms and offering explanations of results \cite{gunning2019darpa}. This is vital to promoting trust and fairness through AI- and ML-driven decision insights. Some traditional ML algorithms, such as decision trees and linear regression, are inherently interpretable but limited in predictive power \cite{carvalho2019machine}. In contrast, while deep learning has achieved exceptional results in many tasks, its complex nature requires XAI techniques to improve the transparency and understanding of its intricate decision making.
Explanation methods can be classified as model-specific and model-agnostic. Model-specific techniques are applied to specific models or groups of models, allowing for an understanding of decisions by investigating their underlying mechanisms, such as explaining coefficient weights in neural networks. Model-agnostic methods, on the other hand, study the relation between input and output variables without having structure of the model, allowing for generalization across multiple models.  The two well-known model-independent explanation strategies in the literature are Local Interpretable Model-Agnostic Explanations (LIME) \cite{ribeiro2016should} and Shapley Additive Explanations (SHAP) \cite{lundberg2017unified}. LIME provides localized explanations by analyzing model predictions with varying input data. The procedure entails generating a new data set consisting of perturbed samples and the corresponding predictions of a black-box model. Alternatively, SHAP is based on the concept of game-theoretically optimal Shapley values as a method to understand the reasoning behind individual predictions. 


\subsection{RRC Signaling Storm Attacks}\label{rrc_signaling}

RRC signaling storms present a significant threat to cellular networks by causing excessive signaling activity that overwhelms the control plane. These storms can occur due to malicious activities, such as attacks by malware or improperly configured applications, or unintentionally from high device density and rapid re-registration attempts. The RRC protocol, particularly in 4G and 5G networks (including those adopting the O-RAN architecture), is vulnerable to such disruptions that compromise network operations~\cite{tabiban2023signaling}.


Detecting RRC signaling storms in real time is essential to enable timely mitigation, especially since response and mitigation strategies differ depending on whether it is a malicious attack or a legitimate high-load scenario. In the event of a malicious attack, targeted actions must be taken to prevent resource waste and protect the network from service degradation. In contrast, during a high-load scenario, the focus is on managing capacity and ensuring that legitimate traffic is prioritized. Differentiating between these two scenarios allows operators to apply appropriate mitigation techniques, preserving network availability, ensuring continuous connectivity, and safeguarding overall network stability and QoS. This proactive approach helps avoid unnecessary resource consumption and ensures that the network performs optimally under both attack and high-load conditions~\cite{kapoor2024signaling}.

\section{Related Work}

Various approaches have been explored in the literature to protect IDSs against such adversarial threats. For example, in \cite{abusnaina2019examining}, the authors propose adversarial training to enhance the detection of adversarial samples. In \cite{nugraha2021detecting}, researchers advocate for a deep learning-based approach specifically tailored for detecting adversaries. However, existing solutions face several challenges. First, many methods require integration during the training phase and direct application within the ML model used by the IDS. This makes the pre-trained models susceptible to analysis by attackers, who can then adapt their tactics to exploit the weaknesses of the model. Moreover, many solutions ignore the specific features of the attack vectors, missing key characteristics. The survey study \cite{neupane2022explainable} highlights the critical importance of explainability in the context of IDS. Although previous studies have made important contributions to the field of IDS by improving explainability, there was no comprehensive framework that combines the strengths of ML robustness against adversarial attacks, real-time applicability, and ensuring the robustness of IDS without human intervention. In our earlier investigation \cite{paltun2024introspective}, we showed that XAI is highly effective in identifying adversaries through the comparison of explanations derived from preprocessed network traffic data with the decisions reached by ML-based IDS. However, this method demands a substantial amount of features for the analysis of adversarial entities, indicating the necessity of a more robust strategy for comprehensive applications. Bridging these gaps with a unified approach has the potential to significantly enhance the efficiency and reliability of IDS systems. To address the above challenges, this paper presents a robust IDS to detect and defend against adversarial attacks incorporating an XAI feature by running the proposed detection model in run time, even with a limited number of features, including white-box and black-box scenarios. We provide the details in the following sections.


\section{Proposed Framework}

\begin{figure}[htbp] 
    \centering
     \includegraphics[width=1\linewidth]{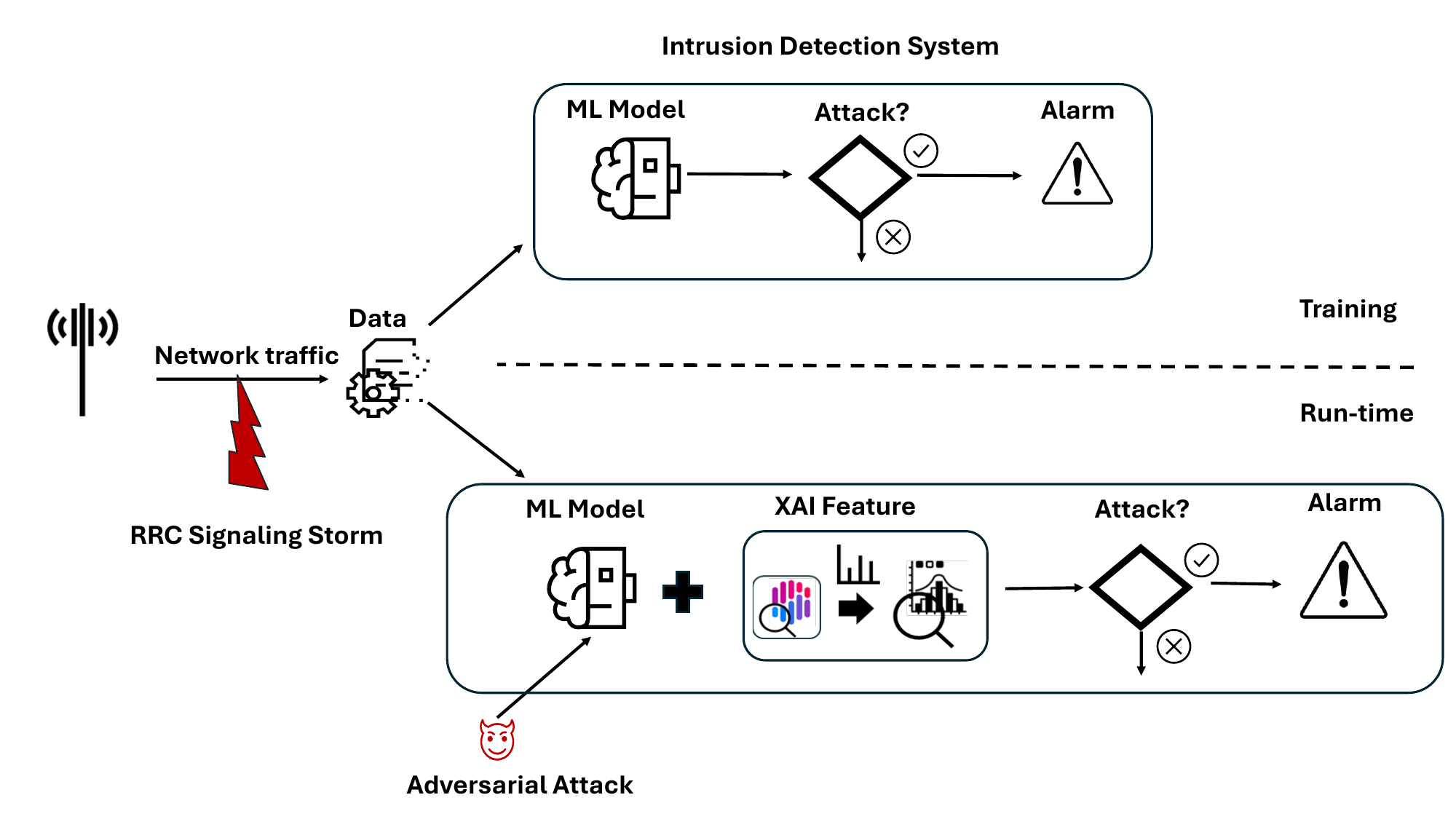}  
    \caption{General structure of the problem and proposed XAI-based adversarial attack detection framework.}
    \label{fig:framework}
\end{figure}

We introduce an agnostic adversarial detection approach to enhance IDS robustness by integrating the XAI feature to perform a real-time assessment. This method is applicable regardless of the IDS model or use case. It aims to strengthen IDS, improving its ability to counteract attackers who use a combination of network and adversarial attacks to bypass detection. The core strategy of the framework involves integrating XAI to identify and highlight features modified by attackers, effectively flagging manipulated network traffic for IDS intervention. Fig.~\ref{fig:framework} presents a general structure of the proposed approach for IDS. Incorporating the XAI feature into the ML model of an IDS assists in determining whether the IDS is vulnerable to adversarial attacks. This is achieved by characterizing the normal behavior pattern of the data through the distribution of SHAP importance values.

\subsection{Adversarial attacks against IDS}\label{attack_ids}

We introduce adversarial instances designed to deceive the IDS. These instances are generated with the best possible results to mislead the IDS by mislabeling malicious traffic as normal. Using techniques such as FGSM \cite{goodfellow2014explaining} and projected gradient descent (PGD) \cite{madry2017towards} can effectively generate adversarial examples. While FGSM is a fast, single-step method that applies a small perturbation in the direction of the gradient of the loss function, PGD is a multistep iterative extension of FGSM that applies successive perturbations and projects the perturbed inputs back onto the valid data domain. In addition to FGSM and PGD, Gaussian attack and Basic Iterative Method (BIM) \cite{kurakin2018adversarial} are other notable techniques for creating adversarial examples. BIM serves as a variation of the PGD approach, iteratively applying small perturbations while also ensuring that each modified input remains within a certain constraint. Meanwhile, the Gaussian attack introduces noise drawn from a Gaussian distribution into the input data. This method can be particularly effective in scenarios where traditional gradient-based approaches may be less successful.

\subsection{Detection of Adversarial attacks}\label{xai_detect}

We propose a novel XAI-based adversarial detection framework designed to determine whether unseen data have been manipulated. The learning process is divided into the training and run-time phases. 
 
\noindent
During the training phase, SHAP feature importance values of training are gathered for each input to characterize the normal behavioral pattern of the data by examining the distribution of these SHAP importance values. 
\begin{itemize}
     \item Let \( X = \{ x_1, x_2, \dots, x_n \} \) be the training data set and let \( f(X) \) be the ML model trained on \( X \). The SHAP importance values for each input \( x_i \) are indicated as \( S(x_i) = \{ S_1(x_i), S_2(x_i), \dots, S_m(x_i) \} \), where \( m \) is the number of features.
    
    \item Let us assume that the distribution of SHAP values for each characteristic \( j \) follows a normal distribution: 
    \[
    S_j(x) \sim \mathcal{N}(\mu_j, \sigma_j^2)
    \]
    where \( \mu_j \) and \( \sigma_j^2 \) are the mean and variance computed from the training data.
 \end{itemize}
\noindent
During run-time, the ML model is used to evaluate unseen data. To assess the behavior of unseen data, we verify if the SHAP feature importance values for this data fall within the normal behavior distribution range of the training data.
 \begin{itemize}
    \item Let \( X' = \{ x'_1, x'_2, \dots, x'_k \} \) be the test data. The SHAP values for a test sample \( x'_i \) are given by \( S(x'_i) = \{ S_1(x'_i), S_2(x'_i), \dots, S_m(x'_i) \} \).
    
    \item The behavior of the test data is verified by checking whether each \( S_j(x'_i) \) falls within the normal behavior range:
    \[
    \mu_j - \lambda \sigma_j \leq S_j(x'_i) \leq \mu_j + \lambda \sigma_j
    \]
    where \( \lambda \) is a threshold parameter (e.g., \( \lambda = 2 \) for a 95\% confidence interval assuming normality).
\end{itemize}
\noindent
If SHAP values of unseen data maintain the same distribution as those of the training data, the performance of the model is considered similar to its training behavior.

\begin{itemize} 
    \item If \( S_j(x'_i) \) falls within this range for all \( j \), the input \( x'_i \) is considered \text{Normal}.
\end{itemize}


\begin{itemize} 
    \item If there exists at least one feature \( j \) such that:
    \[
    S_j(x'_i) < \mu_j - \lambda \sigma_j \quad \text{or} \quad S_j(x'_i) > \mu_j + \lambda \sigma_j
    \]
    then \( x'_i \) is considered \text{Attack}.
\end{itemize}

\section{Experimental Evaluation}

The primary objective of the experimental setup is to show how our proposed approach is capable of detecting and mitigating adversarial instances that cause a significant degradation in the IDS performance which is developed for RRC signaling storm attacks in real-time. We leverage XAI to detect and prevent degradation of IDS performance. The proposed framework seeks to detect potential adversarial attacks by observing significant deviations in the distribution of SHAP importance values for each input in real-time, and to establish a straightforward yet effective zero-touch mitigation strategy.

\subsection{O-RAN Setup}\label{data_colllect}

\begin{figure}[htbp] 
    \centering
     \includegraphics[width=1\linewidth]{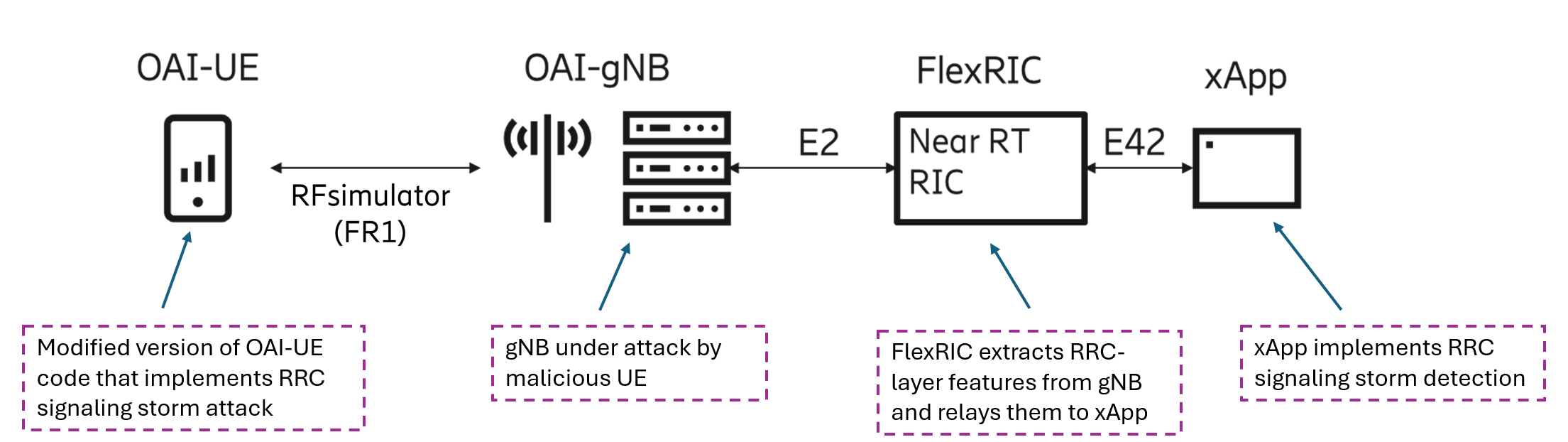}  
    \caption{Steps of xApp feature retrieval from a gNB within OAI setup using FlexRIC.}
    \label{fig:OAI_data}
\end{figure}

The OpenAirInterface (OAI) setup was utilized to implement the RRC signaling attack and extract relevant features for detection. The key components of the setup include OAI-UE (User Equipment), OAI-gNB (a software-based 5G base station), FlexRIC (an open source Near-RT-RIC that extracts RRC-related features from the OAI-gNB and forwards them to an xApp), an xApp (which processes the extracted features to perform detection tasks), and OAI-core. The OAI-UE has been modified to perform a malicious RRC attack by intentionally exploiting vulnerabilities in the RRC signaling process, enabling the generation of abnormal signaling patterns~\cite{kaltenberger2020openairinterface}. The details of data collection are given in Fig. \ref{fig:OAI_data}.

FlexRIC acts as a real-time RAN Intelligent Controller (RIC) platform to enable communication between the xApp and the gNB using the E2 interface. The gNB is configured with an E2 agent that collects RAN metrics and handles control messages.
The xApp subscribes to specific features or metrics (e.g., RRC measurements, CQI, traffic statistics) through FlexRIC. The E2 agent collects the requested features and sends them to FlexRIC via the E2 interface. FlexRIC processes the data received and delivers it to the xApp using its internal communication framework.

\subsection{Experiment Settings}\label{exp_setting}

\subsubsection{Data}
\begin{figure}[htbp]  
    \centering
     \includegraphics[width=1\linewidth]{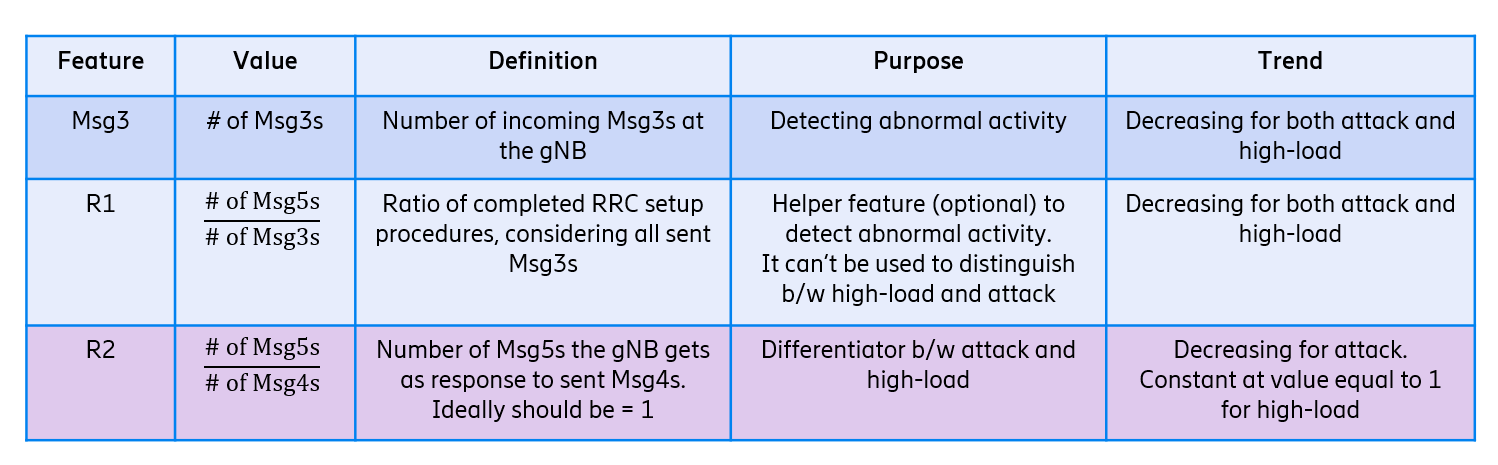}  
    \caption{The key features used for the detection of RRC signaling storms.}
    \label{fig:RRC_features}
\end{figure}

In the experiments, there is 1 malicious User Equipment (UE) and no benign UE. The resource reservation delay by the gNB is about 2.7 seconds, which allocates resources to UE. The Open Air Interface gNB can handle up to 16 resources. The attack rate is around 132 messages per second, exceeding the 90-message-per-second threshold, which overloads the gNB. Consequently, the gNB becomes blocked in approximately 157 milliseconds.For the detection of RRC signaling storms, five key characteristics are collected. Detailed descriptions of these features are given in Fig. \ref{fig:RRC_features}.  A decrease in Msg5s, seen in both attacks and high-load situations, causes $R1$ to drop, which helps to confirm the presence of an anomaly. In an attack, the malicious UE does not respond to Msg4s with Msg5s, while in a high-load scenario, not all UEs will receive Msg4, leading to fewer Msg5s in return. These features, when analyzed together, help create a detection system that identifies signaling storms and distinguishes between attacks and high-load scenarios.

\subsubsection{Intrusion Detection System}\label{sec:ids}

In this paper, IDS is developed using auto-encoders, unsupervised learning neural networks, which are often used to identify anomalies. An auto-encoder comprises an encoder that compresses the input \( \mathbf{x} \in \mathbb{R}^n \) to a latent representation \( \mathbf{z} \in \mathbb{R}^m \) (with \( m < n \)) and a decoder that aims to reconstruct the original data. The key benefit lies in its training on normal data, enabling the capture of typical patterns. During testing, the auto-encoder tries to reconstruct the input based on previous learning. If anomalies exist in test data, the reconstruction error is characterized as

\[
E(\mathbf{x}) = \|\mathbf{x} - \hat{\mathbf{x}}\|
\]
\noindent
(where \( \hat{\mathbf{x}} \) is the reconstructed output), tends to be significantly higher, indicating the existence of unexpected patterns, which can be interpreted as potential anomalies.
\noindent
In this research, the auto-encoder model is trained with a normal traffic data set to ensure it outputs features similar to regular traffic. The similarity between input and output vectors is assessed using vector similarity metrics such as Euclidean distance:

\[
d(\mathbf{x}, \hat{\mathbf{x}}) = \sqrt{\sum_{i=1}^{n} (x_i - \hat{x}_i)^2}
\]
\noindent
If attack traffic features are introduced into the model, we expect to obtain larger distance values compared to those for normal traffic inputs.

\subsubsection{Scenarios}

There are three different experimental setups.

In the initial scenario, we assess the performance of the IDS in two scenarios: one with only an RRC signaling storm attack to disrupt network traffic and another that combined this attack with an adversarial attack on the IDS. Various techniques are used to assess the effectiveness of adversarial attacks.

\begin{itemize}
    \item Initially, we simulate an RRC signaling storm attack characterized by excessive signaling that overloads the control plane. The IDS is then evaluated for its detection performance.
    \item To establish a baseline, IDS performance is evaluated under normal conditions without adversarial attacks. An auto-encoder, described in Section \ref{sec:ids}, is employed for this purpose. 
    \item We present adversarial examples designed to mislead IDS, followed by performance evaluation to determine the level of degradation.
\end{itemize}

In the second scenario, our suggested adversarial attack detection strategy, integrating an XAI feature, aids in determining if the unseen data provided to the IDS falls within the typical behavior distribution of the training data.

\begin{itemize}
    \item During training time, the importance values of each input are calculated using the SHAP method.
    \item Kernel density estimation is used to create a smooth estimate of the distribution from importance values of training data and used to define the normal behavior pattern of the IDS data. 
    \item During run-time, the autoencoder is run to evaluate unseen data. To determine whether a test input is an anomaly based on a distance, we compute a threshold (defined as a Z-Score) based on our known normal behavior (which could be derived from the training data) and then check if the distance of the unseen input exceeds this threshold.
\end{itemize}

Finally, we apply a simple yet effective mitigation approach that observes the predicted attack instances.

\begin{itemize}
    \item If the unseen data follow the same distribution as the training data (i.e. normal), it is considered "Normal".
    \item If an input falls outside the usual distribution range and is classified as an outlier, it is considered manipulated, and we adjust its label accordingly.
\end{itemize}

\subsection{Performance Results}\label{performance}

\begin{figure}[htbp]
    \centering
     \includegraphics[width=1\linewidth]{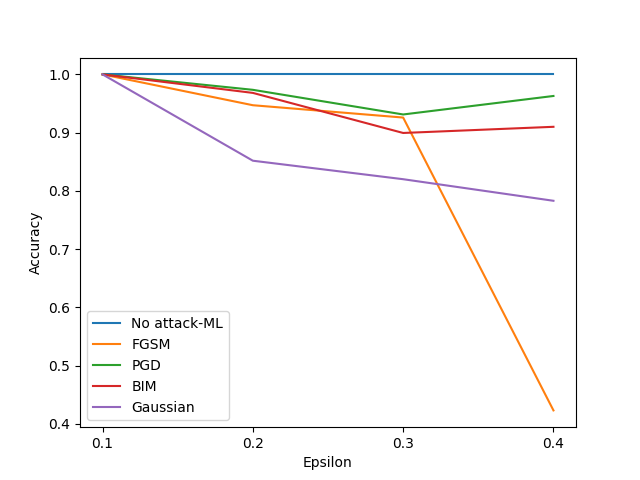}  
    \caption{The accuracy of IDS under different adversarial attack methods.}
    \label{fig:method_comparison}
\end{figure}

Fig.~\ref{fig:method_comparison} shows the accuracy of IDS under different adversarial attack methods. A higher epsilon value indicates stronger perturbations in the X-axis. Y-axis represents the accuracy of the IDS in correctly classifying inputs as either normal or malicious. No attack shows the baseline accuracy of the IDS when there are no adversarial attacks. The accuracy remains constant and high accuracy across all epsilon values indicates the model performs well without adversarial interference. As the epsilon increases, the accuracy drops for the rest of the models, showing that the performance of IDS degrades with stronger adversarial perturbations. The significant drop in accuracy with increasing epsilon suggests that FGSM is highly effective at exploiting vulnerabilities, demonstrating the need for stronger defenses against gradient-based attacks. Although accuracy decreases, the model maintains better performance compared to FGSM, indicating some resilience. This suggests that iterative attacks like PGD are less effective but still significant, highlighting the partial robustness of the model. With slightly better accuracy than PGD, the model shows better handling of iterative attacks, suggesting that existing defenses might be more effective against methods like BIM. The drop in accuracy under Gaussian attack shows high sensitivity to random noise, highlighting the necessity for robust noise features and effective preprocessing.

In the second experiment, to evaluate our proposed approach, we also implemented it with individual feature importance methods such as LIME and permutation importance. Table~\ref{tab:confusion_metrics} shows the comparison of the performance metrics of four approaches that reflect different detection scenarios under the BIM attack. The AE-BIM attack does not apply any adversarial detection method, with a precision of 0.7619 and indicating that it does not detect any false positive, highlighting a challenge in recognizing adversarial examples. The permutation approach shows an improvement with both true negatives and some true positives, indicating better performance in recognizing adversarial perturbations.The LIME method further improves upon Permutation, with a higher accuracy of 0.8307 and a better balance between precision and recall. Finally, the SHAP method demonstrates the best performance across all metrics, achieving a significant accuracy of 0.9259 and an F1-score of 0.8654. In particular, it correctly identifies all positive cases, while maintaining a high precision of 0.7627. This suggests that SHAP is the most effective, providing both high accuracy and robust anomaly detection capabilities.

\begin{table}[h]
    \centering
    \caption{Detection performance of four different methods.}
 \label{tab:confusion_metrics}
    \begin{tabular}{|l|c|c|c|c|}
        \hline
        \textbf{Method} & \textbf{Accuracy} & \textbf{Precision} & \textbf{Recall} & \textbf{F1-Score} \\
        \hline
        AE-BIM Attack & 0.7619 & 0.0000 & 0.0000 & 0.0000 \\
        Permutation & 0.8095 & 0.6364 & 0.4667 & 0.5385 \\
        LIME & 0.8307 & 0.6383 & 0.6667 & 0.6522 \\
        SHAP & 0.9259 & 0.7627 & 1.0000 & 0.8654 \\
        \hline
    \end{tabular}

\end{table}

\begin{figure}[htbp]
    \centering
    \includegraphics[width=1\linewidth]{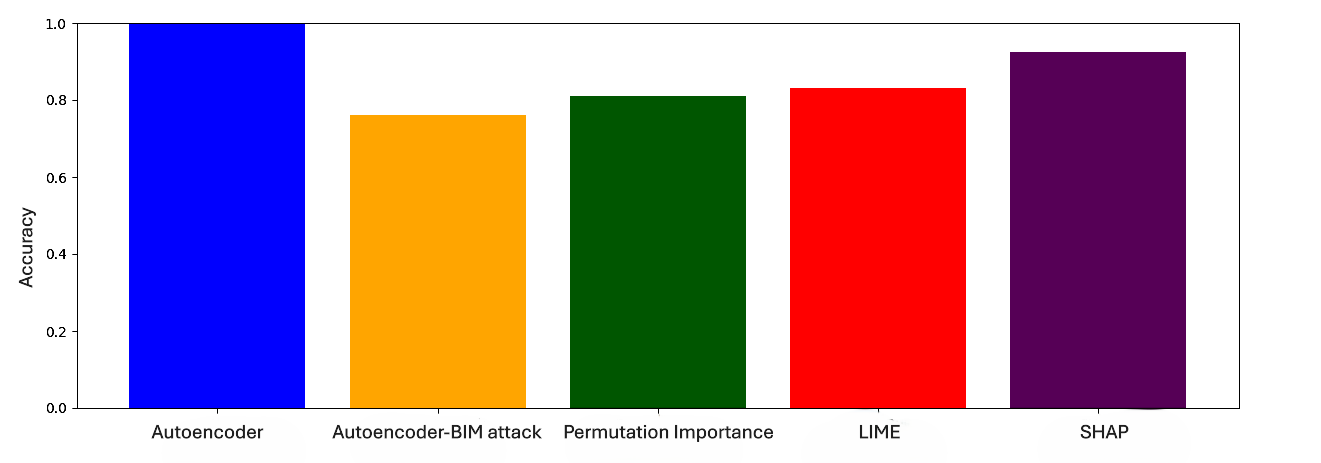}  
    \caption{Method comparison with XAI-based mitigation}
    \label{fig:bar_comparison}
\end{figure}

In the mitigation scenario, Fig.~\ref{fig:bar_comparison} illustrates the mitigation performance of the proposed approach in various scenarios. The permutation importance approach improves accuracy, suggesting detrimental effects from the attack or transformation. Although LIME helps explain predictions, its effectiveness is comparable to that of the permutation method, highlighting its limitations in mitigating attack impacts. In particular, SHAP achieves the highest level of accuracy among the individual feature importance methods, reflecting its superior capability to manage adversarial influences and improve the robustness of the model. These results are consistent with those obtained in earlier experiments.

\section{Conclusion}

In conclusion, the proposed system, which integrates the XAI feature into the adversarial detection method, identifies unusual behaviors indicative of adversarial activities accurately and efficiently. This method demonstrated its effectiveness by improving the performance of IDS in an O-RAN environment, representing notable progress in strengthening cybersecurity measures.The suggested method introduces zero-touch functionality and, when implemented on IDS, provides prompt responses to new threats, significantly reducing potential risks to network infrastructures. Furthermore, with the ongoing evolution of O-RAN, the collaboration of XAI and IDS will be critical to adapt to new vulnerabilities. In the future, ongoing research and development in this area is essential to continuously enhance and refine the efficiency of XAI-enabled cybersecurity measures, ensuring that network defenses remain strong within the constantly changing landscape of cyber threats.

\bibliography{ref.bib}

\begin{thebibliography}{30}

\bibitem{chakraborty2018adversarial}
Anirban Chakraborty, Manaar Alam, Vishal Dey, Anupam Chattopadhyay, and Debdeep
  Mukhopadhyay.
\newblock Adversarial attacks and defences: A survey.
\newblock {\em arXiv preprint arXiv:1810.00069}, 2018.

\bibitem{niknam2022intelligent}
Solmaz Niknam, Abhishek Roy, Harpreet~S Dhillon, Sukhdeep Singh, Rahul Banerji,
  Jeffery~H Reed, Navrati Saxena, and Seungil Yoon.
\newblock Intelligent o-ran for beyond 5g and 6g wireless networks.
\newblock In {\em 2022 IEEE Globecom Workshops (GC Wkshps)}, pages 215--220.
  IEEE, 2022.

  \bibitem{illiano2015detecting}
Vittorio~P Illiano and Emil~C Lupu.
\newblock Detecting malicious data injections in wireless sensor networks: A
  survey.
\newblock {\em ACM Computing Surveys (CSUR)}, 48(2):1--33, 2015.

\bibitem{gunning2019darpa}
David Gunning and David Aha.
\newblock Darpa’s explainable artificial intelligence (xai) program.
\newblock {\em AI magazine}, 40(2):44--58, 2019.

\bibitem{lundberg2017unified}
Scott~M Lundberg and Su-In Lee.
\newblock A unified approach to interpreting model predictions.
\newblock {\em Advances in neural information processing systems}, 30, 2017.

\bibitem{tabiban2023signaling}
Azadeh Tabiban, Hyame Assem Alameddine, Mohammad~A Salahuddin, and Raouf Boutaba.
\newblock Signaling storm in o-ran: Challenges and research opportunities.
\newblock {\em IEEE Communications Magazine}, 2023.

\bibitem{kapoor2024signaling}
Adithya Kapoor, Abiodun Ganiyu, and Vijay~K. Shah.
\newblock Signaling storming attack detection and mitigation in open radio access networks.
\newblock {\em Journal of Student-Scientists' Research}, 6, 2024.

\bibitem{abusnaina2019examining}
Ahmed Abusnaina, Aminollah Khormali, DaeHun Nyang, Murat Yuksel, and Aziz
  Mohaisen.
\newblock Examining the robustness of learning-based ddos detection in software
  defined networks.
\newblock In {\em 2019 IEEE Conference on Dependable and Secure Computing
  (DSC)}, pages 1--8. IEEE, 2019.

\bibitem{nugraha2021detecting}
Beny Nugraha, Naina Kulkarni, and Akash Gopikrishnan.
\newblock Detecting adversarial ddos attacks in software-defined networking
  using deep learning techniques and adversarial training.
\newblock In {\em 2021 IEEE International Conference on Cyber Security and
  Resilience (CSR)}, pages 448--454. IEEE, 2021.

\bibitem{neupane2022explainable}
Subash Neupane, Jesse Ables, William Anderson, Sudip Mittal, Shahram Rahimi,
  Ioana Banicescu, and Maria Seale.
\newblock Explainable intrusion detection systems (x-ids): A survey of current
  methods, challenges, and opportunities.
\newblock {\em IEEE Access}, 10:112392--112415, 2022.


\bibitem{paltun2024introspective}
Bet{\"u}l G{\"u}ven{\c{c}} Paltun and Ramin Fuladi.
\newblock Introspective intrusion detection system through explainable ai.
\newblock In {\em 2024 8th Cyber Security in Networking Conference (CSNet)},
  pages 28--32. IEEE, 2024.

\bibitem{goodfellow2014explaining}
Ian~J Goodfellow, Jonathon Shlens, and Christian Szegedy.
\newblock Explaining and harnessing adversarial examples.
\newblock {\em arXiv preprint arXiv:1412.6572}, 2014.


\bibitem{madry2017towards}
Aleksander Madry, Aleksandar Makelov, Ludwig Schmidt, Dimitris Tsipras, and
  Adrian Vladu.
\newblock Towards deep learning models resistant to adversarial attacks.
\newblock {\em arXiv preprint arXiv:1706.06083}, 2017.


\bibitem{kurakin2018adversarial}
Alexey Kurakin, Ian~J Goodfellow, and Samy Bengio.
\newblock Adversarial examples in the physical world.
\newblock In {\em Artificial intelligence safety and security}, pages 99--112.
  Chapman and Hall/CRC, 2018.

\bibitem{kaltenberger2020openairinterface}
Florian Kaltenberger, Aloizio~P Silva, Abhimanyu Gosain, Luhan Wang, and
  Tien-Thinh Nguyen.
\newblock Openairinterface: Democratizing innovation in the 5g era.
\newblock {\em Computer Networks}, 176:107284, 2020.



\end{thebibliography}
\bibliographystyle{ieeetr}

\end{document}